# Tuning Coherent Light Generation in 2D Semiconductors with Strong Laser Fields

George Trivizas[1], Matthew D. Feinstein[1] and Euclides Almeida[1,2,*]

[1]Department of Physics, Queens College, City University of New York, Flushing, NY 11367, United States of America

[2]The Graduate Center of the City University of New York, New York, NY 10016, United States of America

*Author's email address: euclides.almeida@qc.cuny.edu

**Abstract**

Light generation through optical harmonics plays a pivotal role in photonics, driving innovations in coherent light sources, biological imaging, and spectroscopy. Traditional methods for tuning optical harmonics, including electrostatic gating, are inherently slow, presenting a bottleneck for the performance of integrated photonic devices. While all-optical schemes offer faster tuning, they often rely on resonant excitation, limiting tunability to the relatively long lifetimes of excited electronic states. Here, we demonstrate a novel approach for all-optical control of harmonic generation in two-dimensional semiconductors by using strong, non-resonant laser pulses, with electric field amplitudes approaching $10^9$ V/m. By harnessing the quantum-mechanical coupling of virtual excitons (electron-hole pairs) with photons below resonance, we achieve precise spectral and amplitude control of second- and third-harmonic generation. Additionally, the spectral tunability reaches Terahertz rates, limited by the control beam's pulse duration and multiphoton absorption. These results open new avenues for the development of tunable, compact coherent light sources and ultrafast information processing.

**Introduction**

Since the first observation of second harmonic generation in 1961[1], nonlinear harmonic generation has transformed diverse scientific and technological fields, catalyzing advancements in biological imaging, materials characterization, and laser science. In particular, harmonic generation has facilitated the development of novel coherent light sources by enabling access to spectro-temporal domains, such as the attosecond extreme ultraviolet[2,3], that are otherwise unreachable with conventional laser media. In commercial systems, wavelength tunability of harmonic generation is typically achieved by changing the phase-matching angle through mechanical rotation of nonlinear crystals. Additional tuning methods, such as temperature, strain, or electrical control[4–9] provide some flexibility but are also inherently slow. Electrical gating, while ideal for integrated nonlinear devices, is fundamentally constrained by losses in conventional electronics, thus limiting its operational speed.

All-optical tuning enables much faster operational speeds compared with the traditional methods[10]. Since off-resonant nonlinearities are weak, most optical control methods rely on electronic excitation in resonant materials. This method is constrained by the lifetime of electronic excitations, which in semiconductors spans from a few picoseconds for hot electrons to nanoseconds for the decay to the valence band. A non-resonant all-optical control approach has been demonstrated in second-harmonic generation in two-dimensional $MoS_2$[11]. Amplitude and polarization optical switching is achieved through symmetry considerations of the second-order nonlinearity, and a nearly perfect cancelation of the second harmonic is obtained, albeit with no spectral control.

Here, we demonstrate a method for attaining ultrafast spectral control of harmonic generation using off-resonant strong fields. This is achieved by coupling virtual excitons in two-dimensional semiconductors with photons below resonance from laser pulses with electric field amplitudes exceeding $10^8$ V/m. The energies of the hybrid exciton-photon system formed can be adjusted by the intensity of the control beam, enabling spectral control of second and third-harmonic generation from monolayer transition metal dichalcogenides in the visible range. We attain spectral tuning of up to 26 meV and amplitude modulation of up to 63% for the laser intensities used in our experiments. Moreover, we dynamically tune the third-harmonic spectrum at Terahertz rates, with the tuning speed limited by the gate pulse duration and nonlinear absorption of the control laser pulse. This approach for controlling optical harmonics may pave the way for a novel class of tunable compact light sources, and has broad implications for ultrafast information processing and quantum photonic technologies.

**Results**

Our method for tuning nonlinear generation is based on the AC Stark effect, a quantum mechanical effect first observed by Autler and Townes in a molecular gas[12]. Decades later, the effect was observed in solid-state semiconducting materials[13–16] and, more recently, in two-dimensional transition metal dichalcogenides (TMDCs)[17–20]. These semiconducting materials are known for their extraordinary excitonic properties[21,22], including large tunability and strong optical nonlinearities[9,23,24]. Moreover, their mechanical properties, such as high flexibility and strength[25], favor integration in photonic circuits[26]. The process for tuning third-harmonic generation with strong fields is illustrated in Fig. 1. A fundamental laser pulse is chosen such that the third harmonic signal is generated near the peak of the excitonic frequency of a TMDC material. This pulse is sufficiently weak such that the nonlinear process can be described using perturbation theory, and the macroscopic polarization $P$ is expressed as:

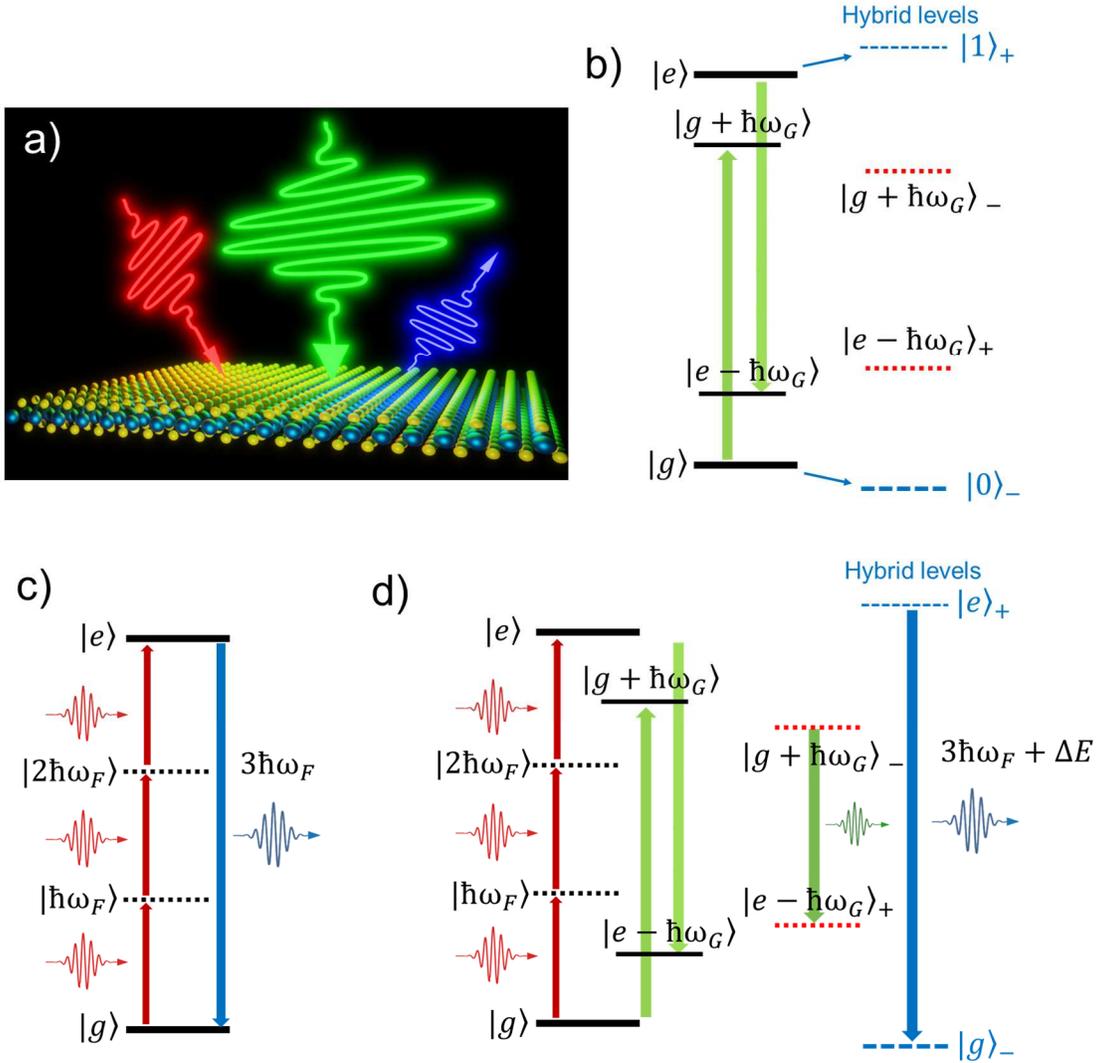

**Figure 1. Tuning harmonic generation in 2D semiconductors using strong fields**. a) Illustration of the harmonic generation process tunable by strong fields. The red (green) arrow represents the fundamental (gate) pulse, and the optical harmonic (blue arrow) is collected in reflection mode. A delay line placed in the path of the gate beam can control the delay between the incident pulses. b) Effect on the exciton energy levels under the presence of the gate pulse. Levels $|g\rangle$ and $|e\rangle$ represent the ground and exciton states respectively, while levels $|g + \hbar\omega_G\rangle$ (absorption) and $|e - \hbar\omega_G\rangle$ (stimulated emission) represent the photon states. The dashed levels represent the hybrid photon-exciton states. c) Energy level diagram for the third harmonic process in the absence of the strong field. d) Third-harmonic process in the presence of the strong field. The third-harmonic signal is generated at $\hbar\omega_{TH} = 3\hbar\omega_F + \Delta E$, where $\Delta E$ is the AC Stark energy shift. Due to conservation of energy, the gate photons are inelastically scattered with energy $\hbar\omega_G - \Delta E$. Our setup is arranged to detect only the third-harmonic photons.

$$P = \chi^{(1)}[\mathcal{E}(\omega_F)] + \chi^{(2)}[\mathcal{E}(\omega_F)]^2 + \chi^{(3)}[\mathcal{E}(\omega_F)]^3 + \cdots \qquad (1)$$

where $\chi^{(n)}$ ($n$ = 1, 2, 3, …) is the n-order nonlinear susceptibility of the material and $\mathcal{E}(\omega_F)$ is the electric field amplitude of the fundamental pulse with carrier frequency $\omega_F$. A strong non-resonant femtosecond laser beam, referred to as the control or the gate beam, overlaps in space and time with the fundamental pulse. The electric field amplitude of the gate beam exceeds $10^8$ V/m, and the nonlinear optical behavior of the system can no longer be treated in the perturbative regime of Eq. 1. The effect of the gate field on the TMDC excitons can be described as the interaction of light with a two-level system with energy splitting $E_0$ (Fig. 1b). Similar to the cavity-exciton interaction[27], the photons of the gate beam will strongly couple with virtual excitons, creating hybrid light-matter states known as photon-dressed virtual excitons[28]. This hybridization leads to energy level repulsion and a blue shift of the dressed exciton spectral line[17,20,28]. For sufficiently weak gate fields, the energy shift $\Delta E$ can be calculated using time-dependent perturbation theory[20], resulting in the functional dependence on the gate field amplitude $|\mathcal{E}(\omega_G)|$ and gate photon energy $\hbar\omega_G$ as:

$$\Delta E = \frac{\mu_0^2 |\mathcal{E}(\omega_G)|^2}{2(E_0 - \hbar\omega_G)} \qquad (2)$$

where $\mu_0$ is the dipole matrix element between the ground and exciton states. The fundamental broadband beam will then "probe" this energy shift through the third harmonic signal. In the absence of the gate field, the third-harmonic is generated at $\hbar\omega_{TH} = 3\hbar\omega_F$ (Fig. 2c). However, applying the gate field will cause a shift of the third harmonic peak to $\hbar\omega_{TH} = 3\hbar\omega_F + \Delta E$ (Fig. 2d). The spectrum of the optical harmonic can be tuned by controlling the intensity of the gate beam and the detuning ($E_0 - \hbar\omega_G$) between the exciton and the gate photons. In our experiments, the gate beam intensity will be the only tuning parameter.

Figure 2 shows the tuning of third-harmonic generation in large-area, CVD-grown monolayer tungsten disulfide ($WS_2$). A broadband, femtosecond fundamental beam from an Optical Parametric Amplifier is selected to be centered at 1,770 nm ($\hbar\omega_F$ = 0.70 eV, FWHM = 80 nm). The central wavelength of the third-harmonic signal is located at $\lambda_{TH}$ = 599.5 nm – slightly above the center of the A-exciton line of the $WS_2$ monolayer (613 nm, $E_0$ = 2.02 eV) – with Full Width at Half Maximum (FWHM) equal to 27.4 nm. The gate beam is obtained from a Ti:Sapphire chirped-pulse laser amplifier, centered at 795 nm ($\hbar\omega_G$ = 1.56 eV) and pulse duration $\tau_G$ = 144 fs. The intensities applied in Fig. 2a – ranging from $I_0$ = 14 to 732 GW/cm$^2$ – correspond to field amplitudes $|\mathcal{E}(\omega_G)|$ from 3.3×10$^8$ to 2.4×10$^9$ V/m. No laser-induced damage to the TMDC was observed at those intensities. In all our measurements, both input beams are linearly polarized along the same direction. As we increase the control laser power, we observe a continuous spectral and amplitude tuning of the third harmonic signal (Fig. 2a). The wavelength tuning is more clearly observed in Fig. 2b, where we normalize the third-harmonic signal for three different control laser intensities. For the highest control laser intensity used ($I_0 = 762$ GW/cm$^2$), the third-harmonic peak shifts by 26 meV (Fig. 2c) and the amplitude decreases to 37% (Fig. 2d) of that of the ungated signal. For gate intensities below $I_0 = 60$ GW/cm$^2$, the inset of Fig. 2c shows that the third-harmonic peak shift increases linearly, as expected from Eq. (2). However, as the gate intensity increases, the energy shift shows a saturation behavior. A sum-frequency generation signal ($\hbar\omega_{SFG} = \hbar\omega_G + \hbar\omega_F$) is observed around 548 nm. While the central frequency of the third-harmonic beam is tunable with the gate intensity, in the sum-frequency signal, only the amplitude changes. This is another signature of the role played by tunable photon-dressed virtual excitons in the third-harmonic generation process. Since the sum-frequency signal is off-resonant, its frequency cannot be tuned by changing the intensity of the gate beam.

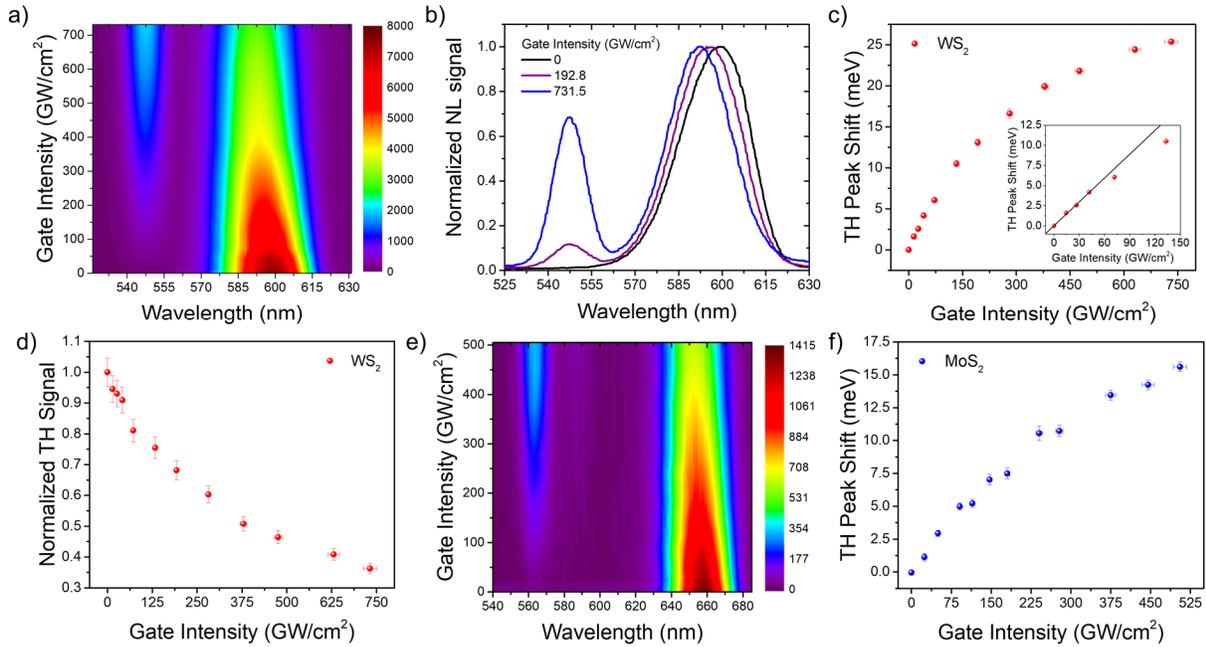

**Figure 2. Experimental results for strong-field tunable third harmonic generation in 2D WS$_2$ and MoS$_2$.** a) Nonlinear generation in WS$_2$ as a function of the gate intensity, showing the third-harmonic signal around 600 nm and the sum-frequency signal at 548 nm. b) Normalized third-harmonic signal. The spectrum of the TH signal is clearly tunable, while the sum-frequency signal remains centered at 548 nm. c) Energy shift of the third harmonic signal as the gate intensity increases. The inset shows a linear fit to the four first data points. d) Integrated amplitude of the TH signal as a function of the gate intensity, normalized to that of the ungated signal. e) Spectrum of the nonlinear signal from monolayer MoS$_2$. The fundamental pulse is tuned to 1,970 nm to address the A-exciton of MoS$_2$. f) Peak shift of the TH signal from monolayer MoS$_2$.

By tuning the fundamental pulse wavelength and changing semiconductor, we attain similar strong field control for third harmonic generation in monolayer MoS$_2$ (Figs. 2e and 2f) as well as for second harmonic generation in monolayer WS$_2$ (Figs. 3a and 3b) and MoS$_2$ (Figs. 3c and 3d), indicating the generality and versatility of our method. In the second harmonic case presented in Fig. 3, we could tune the peak within the 10-20 meV range, and the peak shift grows with the gate intensity, eventually showing a saturation behavior.

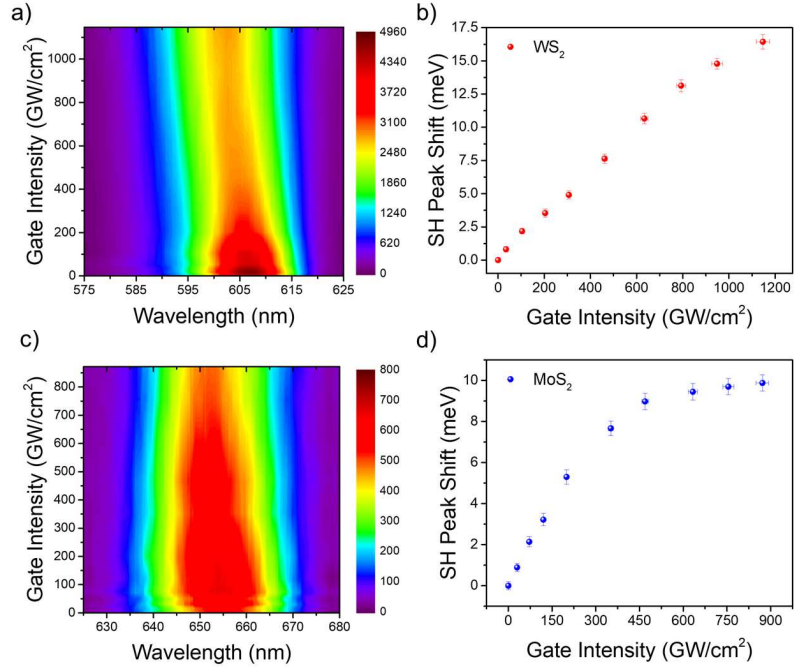

**Figure 3. Tuning second harmonic generation with strong laser fields.** a) Second harmonic from monolayer WS$_2$ and b) peak shift for various gate laser intensities. c) Second harmonic for monolayer MoS$_2$ and d) peak shift.

Since the AC Stark shift is a quantum coherent phenomenon that occurs only during the application of the gate pulse, this allows for fast spectral tuning of the optical harmonic. Figure 4 shows time-resolved spectral measurements of third-harmonic generation in WS$_2$, where we control the delay between the fundamental and gate pulses. The spectrogram of Fig. 4a, measured for a gate beam intensity $I_0$ = 692 GW/cm$^2$, shows spectral and amplitude changes of the third harmonic. The spectrogram also shows a Frequency-Resolved Optical Gating[29] (FROG) trace around $\lambda$ = 548 nm between gate and fundamental beams. At 65 GW/cm$^2$ gate beam intensity (Fig. 4b), the FWHM of the coherent peak of the third-harmonic time-delay profile is 200 fs. This is larger than the convolution of the fundamental and pump pulses calculated from the FROG trace. Notably, as we increase the control laser intensity (Figs. 4c and 4d), the FWHM increases and the slower contribution of the third-harmonic shift becomes more pronounced. This is a signature of nonlinear absorption of the control beam, which causes a shift of the third-harmonic frequency even when the beams no longer temporally overlap. In Fig. 4e we show a measurement for longer delays with $I_0 = 646$ GW/cm$^2$.

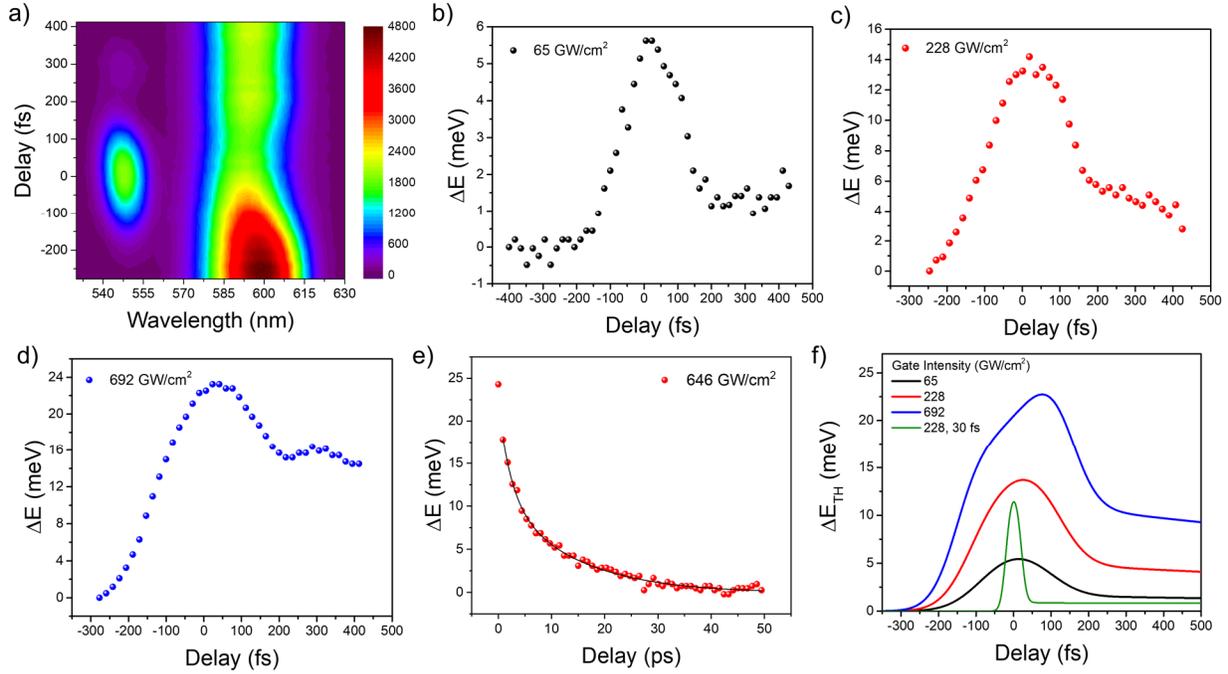

**Figure 4. Dynamic control of third-harmonic generation and phenomenological model.** a) Time-delayed measurements for third-harmonic generation in WS$_2$ for a gate intensity $I_0 = 692$ GW/cm$^2$. The zero-time is set as the maximum of the off-resonant sum-frequency signal. Positive delays indicate that the fundamental pulse lags the gate beam. Energy shift for $I_0 = 65$ GW/cm$^2$ (b), 228 GW/cm$^2$ (c) and 692 GW/cm$^2$ (d). The "echo" signal at $t = 300$ fs is caused by reflection of the gate beam from the optics in the experimental setup. The "echo" also appears in the FROG trace around $\lambda = 548$ nm. e) Decay of the third-harmonic generation signal for longer delays for $I_0 = 646$ GW/cm$^2$ and double exponential fit. The first data point of the fitting was removed to exclude the coherent peak. f) Theoretical curves for the evolution of the third-harmonic signal.

The temporal dynamics of the energy shift is best fit by a sum of two exponential decays with time constants $t_1 = 2.3 \pm 0.5$ ps and $t_2 = 13.6 \pm 1.6$ ps. The picosecond time scale is consistent with reported measurements of hot-carrier relaxation in 2D materials[30]. As the gate beam is nonlinearly absorbed and the conduction band is populated, the exciton line blue shifts, and the ratio between the sharper coherent (AC Stark) peak and the slower component decreases. To capture these dynamics, we construct a phenomenological model based on our experimental observations. The results are shown in Fig. 4f, and details of the model and fitting procedure are given in the Supplementary Information. This model considers the energy shift as the sum of two contributions consisting of coherent ($\Delta E_{ACS}$) and incoherent ($\Delta E_{Inc}$) processes:

$$\Delta E_{TH}(t) = \Delta E_{ACS}(t) + \Delta E_{Inc}(t) \tag{3}$$

The coherent contribution is the dynamic AC Stark effect, instantaneous with the gate pulse intensity $I(t)$. At higher gate intensities, exciton bleaching[31–34] leads to an effective, intensity-dependent transition dipole moment:

$$\mu_{eff}(t) = \frac{\mu_0}{1 + s_1 I(t)} \tag{4}$$

where $s_1$ is a saturation parameter. The incoherent contribution $\Delta E_{Inc}(t)$ is a slower process caused by nonlinear absorption of the gate pulse with two decay constants $t_1$ and $t_2$, where $t_1$ is the only relevant contribution in the time window under consideration. We express the cumulative effects caused by nonlinear absorption by the expression:

$$\Delta E_{Inc}(t) \propto \int_{-\infty}^{t} G(t) e^{-\frac{(t-t')}{t_1}} dt' \tag{5}$$

where $G(t)$ is a function of the gate pulse intensity, and quantifies the exciton frequency shift as the pulse is absorbed by the material. From the behavior of the slow component after the coherent peak, we notice a sublinear growth of the energy shift with the gate intensity. We can express this behavior in $G(t)$ by incorporating saturation effects in the expression:

$$G(t) = \frac{I(t)}{1 + s_2 I(t)} \tag{6}$$

Although the saturation parameters $s_1$ and $s_2$ are not necessarily equal, our fits are simplified by making $s_2 = s_1$. The fitting procedure allowed us to build an approximate expression for the peak shift within the experimental intensity range applied using common fitting parameters. The curves are shown in Fig 4(f). The phenomenological model reproduces a few features of the experimental observations for increasing gate intensities: 1) A saturation of the coherent peak and an increase in its width, 2) a shift of the maximum away from zero delay, and 3) a reduction of the ratio between the coherent and incoherent contributions.

Nonlinear absorption of the gate beam at higher intensities limits the spectral tuning rate to the picosecond regime. The green curve of Fig. 4f shows a prediction based on the model for a laser pulse with $I_0 = 646$ GW/cm$^2$, where the convolution between the gate and fundamental beams was set to 30 fs. The curve is sharply peaked, and the incoherent contribution is reduced, since we deposit less energy for shorter pulses. In the extreme limit, using strong few-cycle light fields, where the field-induced electron transfer from the valence to the conduction band can be greatly reduced[35], will potentially lead to low-loss tuning of coherent light generation in excitonic materials at femtosecond time scales.

**Discussion**

We have proposed and demonstrated a novel scheme for ultrafast tuning of harmonic generation in atomically-thick materials. The scheme involves applying an off-resonant, strong gate beam with electric field amplitudes exceeding $10^8$ V/m, which couples to two-dimensional virtual excitons forming short-lived, hybrid light-matter states (virtual exciton-polaritons). This strong light-matter coupling mechanism enabled fast spectral control of second and third-harmonic generation in $WS_2$ and $MoS_2$, with the time resolution governed by the gate pulse duration and nonlinear absorption of the gate beam. This nonlinear absorption, at the highest gate intensities, caused a large blue shift of the third harmonic. Consequently, the nonlinear signal was composed of a fast, subpicosecond component, and two slow components in the picosecond range. The third-harmonic spectrum was recovered in tens of picoseconds after applying the gate beam. To attain faster recovery, one can use shorter pulses, or gate pulses with lower photon energies to minimize multiphoton absorption. The trade-off for using lower energy photons is a reduction of the AC Stark shift, since the energy shift is inversely proportional to the exciton-photon detuning. At this point, the Bloch-Siegert shift[20,36], which is the coupling with the counter-rotating component of the field and scales as $\Delta E_{BS} \propto (E_0 + \hbar\omega)^{-1}$, will become comparable to the AC Stark shift. Bloch-Siegert shifts as high as 10 meV have been reported in pump-probe experiments in monolayer $WS_2$[20].

Since the present scheme leverages the quantum mechanical coupling between virtual excitons and photons, the principle can be extended to other excitonic materials and nanostructures. While the maximum spectral tuning achieved in the present study in single-layer TMDCs was 26 meV, carefully engineering exciton-photon coupling in TMDCs may improve tunability[37]. The method we presented may also be applied to other exciton-assisted nonlinear processes, such as high-harmonic generation in solids[2,38,39] and multi-wave mixing. These processes cover a wide range of the electromagnetic spectrum, from the Terahertz regime to extreme ultraviolet in thin solids[2]. This may pave the way to novel applications in ultrafast signal processing, integrated light sources and quantum optoelectronics.

**Methods**

*Experiment*

Large area, CVD-grown $WS_2$ and $MoS_2$ monolayers were purchased from 2D Semiconductors, Inc.. The monolayers are supported by a SiO2/Si substrate. The thicknesses of the SiO2 layer and the silicon substrate are 90 nm and 530 μm respectively.

In the all-optical control setup, the fundamental pulses are obtained from an optical parametric amplifier, which generates signal and idler pulses tunable within the 1.04-2.6 μm spectral range. This OPA is optically pumped by an Ti:Sapphire chirped-pulse amplifier (800 nm, 1 kHz). The gate pulse is obtained from the residual beam pumping the OPA. The fundamental pulse energy is set at 50 µJ. The linear polarization of the fundamental, signal and idler beams can be rotated by half-wave plates to ensure that the fundamental and gate pulses are co-polarized. The fundamental beam at 1,770 nm and gate beams Gaussian spot sizes are 90 and 52 µm, respectively.

The TH signal is generated by the sample and collected in reflection mode by the same parabolic mirror, and sent to a CCD-coupled spectrometer via an optical fiber for spectral analysis, or to an electron-multiplying CCD camera for imaging and alignment. Short-pass filters are used to remove stray photons originating from both fundamental and gate pulses.

*Data Analysis and Processing*

We applied the median filtering method (10-point window) to remove cosmic rays from the spectrum of the nonlinear signal acquired with a 1024x256 pixels$^2$ cryogenically-cooled CCD camera. The peak wavelength was determined by fitting an asymmetric double sigmoid function to the median-filtered nonlinear signal using the least squares method. The y-axis error bars correspond to the sum of the error given by the standard deviation of the fitting and the error caused by the gate laser fluctuation.

**Acknowledgements**

Research was sponsored by the Air Force Office of Scientific Research and was accomplished under Grant Number W911NF-23-1-0156. The views and conclusions contained in this document are those of the authors and should not be interpreted as representing the official policies, either expressed or implied, of the Air Force Office of Scientific Research or the U.S. Government. The U.S. Government is authorized to reproduce and distribute reprints for Government purposes notwithstanding any copyright notation herein. Support for this project was provided by a PSC-CUNY Award, jointly funded by The Professional Staff Congress and The City University of New York. The authors acknowledge fruitful scientific discussions with Vinod Menon and Igor Kuskovsky.

**Author Contribution**

EA conceived the idea and led the project. All authors performed the experiments, discussed the results and wrote the manuscript. GT and EA performed the data processing and analysis. EA formulated the phenomenological model.

**Data Availability**

Data supporting this work is available upon request to the corresponding author.

**Conflict of Interests**

The authors declare no conflict of interest.

**References**

1. Franken, P. A., Hill, A. E., Peters, C. W. & Weinreich, G. Generation of Optical Harmonics. *Phys Rev Lett* **7**, 118–119 (1961).

2. Ghimire, S. & Reis, D. A. High-harmonic generation from solids. *Nat Phys* **15**, 10–16 (2019).

3. Krausz, F. & Ivanov, M. Attosecond physics. *Rev Mod Phys* **81**, 163–234 (2009).

4. Seyler, K. L. *et al.* Electrical control of second-harmonic generation in a WSe2 monolayer transistor. *Nat Nanotechnol* **10**, 407–411 (2015).

5. Ren, M.-L., Berger, J. S., Liu, W., Liu, G. & Agarwal, R. Strong modulation of second-harmonic generation with very large contrast in semiconducting CdS via high-field domain. *Nat Commun* **9**, 186–188 (2018).

6. Wang, Y. *et al.* Direct electrical modulation of second-order optical susceptibility via phase transitions. *Nat Electron* **4**, 725–730 (2021).

7. Soavi, G. *et al.* Broadband, electrically tunable third-harmonic generation in graphene. *Nat Nanotechnol* **13**, 583–588 (2018).

8. Feinstein, M. D., Andronikides, A. & Almeida, E. Electro-Optical Switching in a Nonlinear Metasurface. (2024).

9. Zhou, R., Krasnok, A., Hussain, N., Yang, S. & Ullah, K. Controlling the harmonic generation in transition metal dichalcogenides and their heterostructures. *Nanophotonics* **11**, 3007–3034 (2022).

10. Ossiander, M. *et al.* The speed limit of optoelectronics. *Nat Commun* **13**, 1620 (2022).

11. Klimmer, S. *et al.* All-optical polarization and amplitude modulation of second-harmonic generation in atomically thin semiconductors. *Nat Photonics* **15**, 837–842 (2021).

12. Autler, S. H. & Townes, C. H. Stark Effect in Rapidly Varying Fields. *Physical review* **100**, 703–722 (1955).

13. LEHMEN, A. V. O. N., CHEMLA, D. S., ZUCKER, J. E. & HERITAGE, J. P. Optical Stark effect on excitons in GaAs quantum wells. *Opt Lett* **11**, 609–611 (1986).

14. Mysyrowicz, A. *et al.* 'Dressed Excitons' in a Multiple-Quantum-Well Structure: Evidence for an Optical Stark Effect with Femtosecond Response Time. *Phys Rev Lett* **56**, 2748–2751 (1986).

15. Unold, T., Mueller, K., Lienau, C., Elsaesser, T. & Wieck, A. D. Optical Stark Effect in a Quantum Dot: Ultrafast Control of Single Exciton Polarizations. *Phys Rev Lett* **92**, 157401 (2004).

16. Yang, Y. *et al.* Large polarization-dependent exciton optical Stark effect in lead iodide perovskites. *Nat Commun* **7**, 12613 (2016).

17. Sie, E. J. *et al.* Valley-selective optical Stark effect in monolayer WS2. *Nat Mater* **14**, 290–294 (2015).

18. Cunningham, P. D., Hanbicki, A. T., Reinecke, T. L., McCreary, K. M. & Jonker, B. T. Resonant optical Stark effect in monolayer WS2. *Nat Commun* **10**, 1–8 (2019).

19. Kim, J. *et al.* Ultrafast generation of pseudo-magnetic field for valley excitons in $WSe_2$ monolayers. *Science (American Association for the Advancement of Science)* **346**, 1205–1208 (2014).

20. Sie, E. J. *et al.* Large, valley-exclusive Bloch-Siegert shift in monolayer WS2. *Science (1979)* **355**, 1066–1069 (2017).

21. Wang, G. *et al.* Colloquium: Excitons in atomically thin transition metal dichalcogenides. *Rev Mod Phys* **90**, 21001 (2018).

22. Mueller, T. & Malic, E. Exciton physics and device application of two-dimensional transition metal dichalcogenide semiconductors. *NPJ 2D Mater Appl* **2**, 29 (2018).

23. Malard, L. M., Alencar, T. V., Barboza, A. P. M., Mak, K. F. & de Paula, A. M. Observation of intense second harmonic generation from $MoS_2$ atomic crystals. *Phys Rev B* **87**, 201401 (2013).

24. Wen, X., Gong, Z. & Li, D. Nonlinear optics of two-dimensional transition metal dichalcogenides. *InfoMat* **1**, 317–337 (2019).

25. Akinwande, D. *et al.* A review on mechanics and mechanical properties of 2D materials—Graphene and beyond. *Extreme Mech Lett* **13**, 42–77 (2017).

26. Datta, I. *et al.* Low-loss composite photonic platform based on 2D semiconductor monolayers. *Nat Photonics* **14**, 256–262 (2020).

27. Liu, X. *et al.* Strong light–matter coupling in two-dimensional atomic crystals. *Nat Photonics* **9**, 30–34 (2015).

28. Mysyrowicz, A. *et al.* 'Dressed Excitons' in a Multiple-Quantum-Well Structure: Evidence for an Optical Stark Effect with Femtosecond Response Time. *Phys Rev Lett* **56**, 2748–2751 (1986).

29. Trebino, R. *et al.* Measuring ultrashort laser pulses in the time-frequency domain using frequency-resolved optical gating. *Review of Scientific Instruments* **68**, 3277–3295 (1997).

30. Ruppert, C., Chernikov, A., Hill, H. M., Rigosi, A. F. & Heinz, T. F. The Role of Electronic and Phononic Excitation in the Optical Response of Monolayer WS2 after Ultrafast Excitation. *Nano Lett* **17**, 644–651 (2017).

31. Katsuyama, T. & Ogawa, K. Dynamic response of exciton absorption in GaAs/AlGaAs quantum wells under high power excitation. *Opt Commun* **71**, 351–354 (1989).

32. Knox, W. H., Chemla, D. S., Miller, D. A. B., Stark, J. B. & Schmitt-Rink, S. Femtosecond ac Stark effect in semiconductor quantum wells: Extreme low- and high-intensity limits. *Phys Rev Lett* **62**, 1189–1192 (1989).

33. Becker, P. C., Fork, R. L., Brito Cruz, C. H., Gordon, J. P. & Shank, C. V. Optical Stark Effect in Organic Dyes Probed with Optical Pulses of 6-fs Duration. *Phys Rev Lett* **60**, 2462–2464 (1988).

34. JOFFRE, M., HULIN, D. & ANTONETTI, A. FEMTOSECOND STUDY OF THE OPTICAL STARK EFFECT IN MULTIPLE QUANTUM WELL STRUCTURES. *Le Journal de Physique Colloques* **48**, C5-537-C5-540 (1987).

35. Sommer, A. *et al.* Attosecond nonlinear polarization and light–matter energy transfer in solids. *Nature* **534**, 86–90 (2016).

36. Bloch, F. & Siegert, A. Magnetic Resonance for Nonrotating Fields. *Physical review* **57**, 522–527 (1940).

37. Schneider, C., Glazov, M. M., Korn, T., Höfling, S. & Urbaszek, B. Two-dimensional semiconductors in the regime of strong light-matter coupling. *Nat Commun* **9**, 2695 (2018).

38. Liu, H. *et al.* High-harmonic generation from an atomically thin semiconductor. *Nat Phys* **13**, 262–265 (2017).

39. Yoshikawa, N. *et al.* Interband resonant high-harmonic generation by valley polarized electron–hole pairs. *Nat Commun* **10**, 1–7 (2019).